\documentclass{emulateapj}
\usepackage{graphicx}
\providecommand{\tabularnewline}{\\}

\shorttitle{Clearing the Dust from Globular Clusters}
\shortauthors{Umbreit, Chatterjee \& Rasio}

\begin{document}

\title{Clearing the Dust from Globular Clusters}

\author{Stefan Umbreit,\altaffilmark{1} Sourav Chatterjee, 
\altaffilmark{1} and
Frederic A.\ Rasio\altaffilmark{1}}

\affil{$^{1}$ Department of Physics and Astronomy, Northwestern  
University,
2131 Tech Drive, Evanston, IL, 60208}
\email{s-umbreit@northwestern.edu}

\begin{abstract}
Recent \emph{Spitzer} observations of the globular cluster M15
detected dust associated with its intracluster medium.
Surprisingly, these observations imply that the dust must be very  
short-lived
compared to the time since the last Galactic plane crossing of the
cluster. Here we propose
a simple mechanism to explain this short lifetime. We argue that the  
kinetic
energy of the material
ejected during a stellar collision may be sufficient to remove the
gas and dust entirely from a cluster,
or to remove the gas as a wind, in addition to partially destroying the dust.
By calculating the
rate of stellar collisions using an $N$-body model for the cluster,
we find remarkable agreement between the average time between collisions
and the inferred dust lifetime in this cluster, suggesting a possible  
close
relation between the two phenomena. Furthermore, we also obtain
the birthrate of blue stragglers formed through collisions in M15. By  
comparing
with the observed number of blue stragglers, we derive an upper limit
for their average lifetime which turns out to be consistent with  
recent model
calculations, thereby lending further support to our model.
\end{abstract}


\section{Introduction}

\nocite{*}In recent years, infrared space telescopes like \emph{IRAS,
ISO} and especially \emph{Spitzer} made it possible to investigate
not only the stellar content of globular clusters (GCs) but also their
intracluster medium (ICM), consisting of gas and dust lost by giants.  
Evans
et al.\ (2003) were the first to clearly detect thermal dust emission
from the core of the GC M15 (NGC~7078) with {\em ISO}. They found a dust
mass an order of magnitude lower than one would expect based
on the number, mass-loss rates and lifetimes of horizontal branch
(HB) stars, even at the extremely low metallicity of M15. Subsequent,
more sensitive measurements with {\em Spitzer}
resulted in a dust mass of $9\times10^{-4} 
\,{\rm M}_{\odot}$ indicating an accumulation time-span not much longer than 
$10^{6}\,$yr (Boyer et al.\ 2006). While such a short
time is commonly explained by ram-pressure stripping during Galactic
plane passages, the time since the last passage for M15 is about
$4\times10^{7}\,$yr, an order of magnitude larger. Therefore,
additional processes have been suggested to explain short ICM
lifetimes, including blow-out of nova explosions, fast winds from
stars, relativistic winds from millisecond pulsars
(Spergel 1991), and radiative ejection by the strong radiation
field in M15 (Smith 1999).

Another possibility that has not
yet been investigated is related to the outcome of stellar collisions.
 From hydrodynamic simulations it is known that colliding main-sequence
stars (MSS) lose typically $\sim1\%$ of their mass 
(see, e.g., Lombardi et al.\ 1996).
As this material is ejected with speeds typically a few
times the escape speed from the stellar surface, it will transfer a
significant amount of energy to the ICM and possibly reduce the  
amount of observable dust considerably. For instance, if the gas, released
through a collision between two $0.5\,{\rm M}_{\odot}$ MSSs,
has a mass of $0.02\,{\rm M}_{\odot}$ and leaves with twice the
escape velocity ($\approx1200\,{\rm km\, s}^{-1}$), it possesses
enough kinetic energy to accelerate more than $15\,{\rm M}_{\odot}$,
the estimated total mass of ICM gas in M15 (Boyer et al.\ 2006), up to a
speed of $40\,{\rm km\, s}^{-1}$, the escape speed from the center of  
M15.

The inferred dust lifetime in the ICM would then require that
such a stellar collision happens every $\sim 1\,$Myr. Given that M15
has a central density of $n\gtrsim10^{6}\,{\rm M}_{\odot}\,{\rm pc}^{-3}$
and a velocity dispersion of $\sigma\approx10\,{\rm km}\,{\rm s}^{-1}$ 
(see, e.g., McNamara et al.\ 2004), the collision time $T_{\rm coll}=
7\times10^{11}\,{\rm yr}\,\left(\frac{10^{5}{\rm pc}^{-3}}{\nu}\right) 
\,\left(\frac{\sigma}{100\,{\rm km}\,{\rm s}^{-1}}\right)$
for a $1\,{\rm M}_{\odot}$ MSS is $\sim 10^{10}\,{\rm yr}$ (Binney \&  
Tremaine 1987).
This results in at least one collision every $<5\,{\rm Myr}$ for the  
more
than $2000$ stars in the core (e.g., Dull et al.\ 1997). From this
rough estimate one can already see that this mechanism
is indeed promising.

However, there are several factors in M15 that complicate
a better estimate of the collision time. Firstly, in GCs as old
as M15, all MSSs have masses below $0.8\,{\rm M}_{\odot}$, while initially
more massive stars have produced remnants like white dwarfs
(WDs) and neutron stars (NSs), which will have different interaction  
rates.
Secondly, because of mass segregation the lower-mass MSSs
may concentrate in lower density regions, while more massive
WDs and NSs concentrate near the center. Finally,
it is not {\em a priori} clear that enough energy from the
gas ejected in a stellar collision can be effectively transferred to  
the ICM
since losses through radiative shocks may be important.

We address the issue of mass-segregation considering a specific
model of M15 given in McNamara et al.\ (2004), which is based on an
$N$-body model of Baumgardt et al.\ (2003). This
allows us to estimate collision times involving different stellar 
populations in the cluster. Furthermore, we obtain a minimum formation 
rate for blue stragglers (BSs) through MS-MS mergers. Using 
the number of observed BSs in the core of M15 we estimate an average 
BS lifetime and compare it with evolutionary models, as an extra 
check on our basic model (\S\ref {sec:Main-Sequence-Collision-Time}).
In \S\ref{gas_ICM} we discuss how the ejected
material from a stellar collision interacts with the ICM and how it can
clear signatures of dust emission. We conclude in
\S\ref{conclusion}.

\section{Collision Times}
\label{sec:Main-Sequence-Collision-Time}

We define a collision between two stars with radii $R_{1}$ and $R_{2}$
to occur whenever their distance $d\leq R_{1}+R_{2}$. The
average local collision time $T_{\rm coll}^{(1,2)}(r)$ for one star of
type {}``1'' to collide with a star of type {}``2'' can be written
as (compare to Binney \& Tremaine 1987 their eq. 8-122)
\begin{eqnarray}
\frac{1}{T_{\rm coll}^{(1,2)}(r)} & = & 4\sqrt{\pi}n_{2}(r)\sigma(r)\label{eq:local-coll-time}\\
  &  & \times\left(\left(R_{1}+R_{2}\right)^{2}+
  \frac{G\left(M_{1}+M_{2}\right)\left(R_{1}+R_{2} 
\right)}{2\sigma(r)^{2}}\right)\,.\nonumber 
\end{eqnarray}
where $n_{1,2}$ is the number density of field stars {}``2'' and stars
of type {}``1'', $M_{1,2}$
their masses, and $\sigma$ the one-dimensional
velocity dispersion (assuming a Maxwellian velocity distribution for
both species with 
$\sigma(r)=\sqrt{(\sigma_1(r)^2+\sigma_2(r)^2)/2}$).
To get the total number of collisions per unit time in the
cluster between these two species, eq.~\ref{eq:local-coll-time}
is integrated over the whole cluster,
\begin{equation}
\frac{dN_{coll,tot}}{dt}=4\pi\int_{0}^{\infty}dr\, r^{2}\, n_{1}(r)\, 
\frac{1}{T_{coll}(r)}\label{eq:totat-coll-rate}
\end{equation}
where $r$ is the radial position in the cluster.

In order to account for
a continuous stellar mass spectrum and the mass-radius relationship of
stars, we take local averages $R_{1,2}(r)$ and $M_{1,2}(r)$
at position $r$. 
We find that in the N-body model for M15 these profiles, as well as 
$n_{1,2}(r)$, can be well represented
by power-laws over a sufficiently large range of 
$0.025\,{\rm pc}<r<2\,{\rm pc}$. 
Inside of $0.025\, {\rm pc}$
there are almost no MSSs, while outside of $\approx1\,{\rm pc}$ the
contribution of the integrand in eq.~\ref{eq:totat-coll-rate} becomes
rapidly negligible. We do not consider collisions between MSSs
and NSs as the NS retention fraction in GCs is expected to be low  
(Pfahl \& Rappaport 2001; Dull et al. 1997). We also do not consider 
collisions between giants and MSSs
as the escape velocity at the surface of a giant, and even more so the
expected energy of the ejected material, is an order of magnitude
lower than for MSSs. For simplicity we choose a constant $\sigma=11\, 
{\rm km\, s}^{-1}$, as $\sigma$ does not vary much within 
$1\, {\rm pc}$ ($\pm 1\, {\rm km\, s}^{-1}$; compare Dull et al. 1997) 
and also agrees with the value obtained by McNamara \& Baumgardt (2004) for M15 
for a similar region $(r<0.3'\approx 0.8\rm{ pc})$.
The slight variations within $1\, {\rm pc}$ are 
accounted for in the error estimates of the collision rates.  
With all quantities given as power-laws, we solve 
eq.~\ref{eq:totat-coll-rate}
analytically. Our calculations are based on the results of
$N$-body simulations by Baumgardt et al.\ (2003), which were
scaled to fit the velocity dispersion profile of M15 in McNamara et
al.\ (2004). Their model consisted of initially $130,072$ stars with
a realistic mass spectrum, and included a treatment of stellar  
evolution and
the Galactic tidal field. They also take into account velocity kicks  
imparted to
NSs at birth and consider two extreme cases, one where all
NSs are retained, and one where all NSs are removed from the cluster.
Here we only consider the latter case, since, as mentioned before, 
the actual NS retention fraction is expected to be very low.

In Table~\ref{tab:Parameters-fits} 
\begin{table}
\begin{center}
\caption{\label{tab:Parameters-fits}Parameters of the power-law fits for
the mass ($M$), density ($n$), and stellar radius ($R$) for
WDs and MSSs. The power-laws are of
the form $a=a_{0}\,(r/r_{0})^{b}$ with $r_{0}=1\,{\rm pc}$.}
\begin{tabular}{|c|c|c|}
\hline
&
$a_{0}$&
$b$\tabularnewline
\hline
\hline
$n_{MS}$&
$(3.9\pm0.2)\times10^{3}{\rm pc}^{-3}$&
$-1.620\pm0.006$\tabularnewline
\hline
$n_{WD}$&
$(3.3\pm0.1)\times10^{3}{\rm pc}^{-3}$&
$-2.280\pm0.009$\tabularnewline
\hline
$R_{MS}$&
$(0.51\pm0.05)\,{\rm R}_{\odot}$&
$-0.143\pm0.008$\tabularnewline
\hline
$R^2_{MS}$&
$(0.32\pm0.05)\,{\rm R}_{\odot}$&
$-0.26\pm0.01$\tabularnewline
\hline
$M_{MS}$&
$(0.49\pm0.02)\,{\rm M}_{\odot}$&
$-0.105\pm0.002$\tabularnewline
\hline
$M_{WD}$&
$(0.74\pm0.02)\,{\rm M}_{\odot}$&
$-0.117\pm0.002$\tabularnewline
\hline
\end{tabular}
\end{center}
\end{table}
the fit parameters for $n$, $R$, $R^2$
and $M$ for WDs and MSSs in M15 are shown. As can
be seen, through mass segregation, MSSs
are more abundant outside $1\,{\rm pc}$ while the more massive WDs
dominate the central $0.75\,{\rm pc}$ where collisions are most likely
to happen. As a consequence, we should expect more collisions between
WDs and MSSs than between MSSs. From eq.~\ref{eq:totat-coll-rate}, we obtain
$dN_{coll,MS-WD}/dt=(2\pm1)\times10^{-7}\,{\rm yr}^{-1}$
and $dN_{coll,MS-MS}/dt=(4\pm1)\times10^{-8}\,{\rm yr}^{-1}$, where
$N_{coll,X-Y}$ is the number of collisions between $X$ and $Y$. It
follows that, given the current state of M15, we expect one
collision every $(5\pm2)\times10^{6}\,{\rm yr}$ between a WD and
a MSS and every $(3\pm1)\times10^{7}\,{\rm yr}$ a collision between
two MSSs, resulting in one collision every $(4\pm2)\times10^{6} 
\,{\rm yr}$ that releases high-velocity gas into the ICM.

It is rather remarkable how closely this timescale
coincides with the estimated ICM dust lifetime,
suggesting a direct connection between dust clearing and
stellar collisions.

%Even considering only collisions between MSSs,
%the resulting collision timescale is much smaller, at least by
%a factor of $5$, than the time since the last Galactic-plane passage.

\begin{figure}
\begin{center}
\plotone{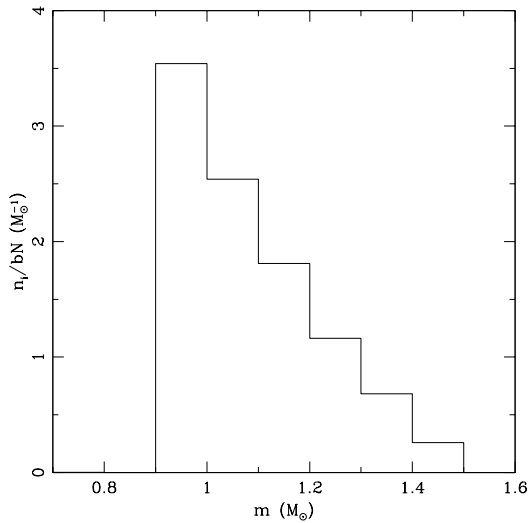}
\caption{Estimated mass distribution of BS candidates created via
collisions between two MSSs with combined mass $>0.8 \rm{M}_{\odot}$.
$n_{i}$, $b$, and $N$ are the number of BSs in the $i$th bin, the bin
size, and the total number of BSs, respectively.}
\label{fig:bs_mass}   
\end{center}
\end{figure}

Using a
similar approach, we estimate the BS formation rate, defining BSs
as two merged MSSs with a combined mass $>1\,{\rm M}_{\odot}$,  
significantly
larger than the MS turn-off mass ($0.8\,{\rm M}_{\odot}$).  
After binning
all MSSs into $0.1\,{\rm M}_{\odot}$ bins, we determine the collision
rate between all mass bins with combined mass $>1\,{\rm M}_{\odot}$. This also
allows us to determine the expected mass spectrum for BSs for comparison
with future observations.
We obtain a total collision rate of $(6.4\pm0.7)\times10^{-3}\,{\rm  
Myr}^{-1}$
which, together with the $6-7$ BSs observed in M15, implies
an average BS lifetime of about $1\,{\rm Gyr}$. This is also  
consistent with recent BS
evolution models (e.g., Leigh et al.\ 2007; Sills et al.\ 2001). 
However, we note that this is certainly an upper limit, given that
the presence of binaries would increase the formation rate through 
binary mergers and resonant interactions 
(Fregeau et al.\ 2004; Mapelli et al.\ 2004).  
For example, for a Plummer sphere with a Kroupa mass function ranging
from $0.3$ to $3.0\,\rm{M_\odot}$ and a binary fraction of $30\%$, the
rate of collisions mediated by binary-single and binary-binary interactions 
can be $\sim 1$ order of magnitude larger
than from single-single interactions (Chatterjee et al.\ 2008). In Figure~\ref{fig:bs_mass} 
the mass distribution for BSs is shown. As expected 
it decreases for higher mass BSs, since the number 
density for the lower mass MSSs is much higher compared to that for the close 
to turn-off MSSs.

\section{Interaction of Collision Ejecta with the ICM}
\label{gas_ICM}

Interactions between very fast moving gas and a low-density interstellar
ambient medium have been extensively investigated in the context of
supernova remnants. Their evolution progresses in several distinct  
stages
(Chevalier 1977): the {}``ejecta dominated'' (ED) stage, 
characterized by a freely expanding blast-shock wave until the mass
of the swept up material is comparable to the mass of the ejecta,
the {}``Sedov-Taylor'' (ST) stage, where the blast-wave expands
adiabatically, and the {}``pressure-driven snowplow'' (PDS) stage,
where a thin shell forms which {}``snowplows'' through the ambient
medium, driven by the pressure of the hot interior in addition to
its own momentum (Cox 1972; Cioffi et al.\ 1988). 
Cioffi et al.\  (1988) obtained a simple offset power-law solution that
describes the kinematics of the blast wave in the PDS stage,  
which can be written as $v_{s}=v_{PDS}\left(R_{s}/R_{PDS}\right)^{-7/3}$, where
$v_{s}$ and $R_{s}$ are the velocity and radial position of the
shock front, respectively, relative to the site of the collision, and
$v_{PDS}$ and $R_{PDS}$ are the velocity and radial position of
the shock front at the transition from the ST stage to the PDS stage.  
The transition values are given by 
$v_{PDS}=413n_{0}^{1/7}\zeta^{3/14}_{m}E_{51}^{1/14}\,{\rm km\, s}^{-1}$
and $R_{PDS}=14.0\, E_{51}^{2/7}\zeta^{-1/7}_{m}n_{0}^{-3/7}
\,{\rm pc}$, where $E_{51}$
is the kinetic energy of the ejected material in units of $10^{51}\, 
{\rm ergs}$, $n_{0}$ the density of the ambient medium in 
${\rm cm}^{-3}$, and $\zeta_m$ is the heavy element abundance relative to
solar abundances.
An interesting property of the flow of the post-shock gas is that in the 
PDS stage its mean velocity has the same value as the shock speed 
(Cioffi et al.\ 1988).
For our problem this means that, provided $v_{s}$ is larger than the 
cluster escape speed after the 
shock swept up all of the ICM, the shocked gas can entirely escape the 
cluster.

It is now interesting to see if such an ICM removal mechanism might be
applicable to M15. For this estimate we assume an ICM 
mass of $15\,{\rm M}_{\odot}$, which is expected given the amount of dust
detected by Boyer et al.\  (2006) and the low metallicity of M15
($\zeta_m=10^{-2.4}$).
In order to estimate $E_{51}$ for collisions between MSS 
we can use the results of Lombardi et al.\ (2002) (their Table 3),
obtaining $E_{51}=1-11\times10^{-4}$. Unfortunately, there are no similar
results for MS--WD collisions. However, given that the mass loss
is presumably between $15\%$ to $50\%$ (Ruffert \& Mueller
1991; Ruffert 1992) and that the velocity of the ejecta should be of the
order of the escape speed of the MSS ($\approx500\, {\rm km\, s}^{-1}$)
we obtain a similar range for $E_{51}$.  Note that,
although the ejecta are ejected nearly isotropically 
(Lombardi et al.\ 1996), the ICM is likely to have a more irregular 
structure, as the patchy dust emission in M15 indicates. Thus, only a 
fraction of the released energy might actually be transferred to the ICM. 
As the 3D structure of the ICM is unknown, we limit our analysis to a 
fiducial ejecta energy of $E_{51}= 3\times10^{-4}$ but also determine the 
minimum $E_{51}$ value to remove all of the ICM gas from the cluster, noting 
that these values are to be understood as {}``effective'' energies ramming 
into the ICM gas.
Similarly, estimates for the ICM density are also rather uncertain. 
If we assume that all the gas is contained within a radius of 
$1-2\, \rm{pc}$ (the approximate radial position of the dust emission), 
the density of a $15\, \rm{M}_\odot$ gas cloud would be between 
$20-150\, \rm{cm}^{-3}$. 
For simplicity, we also assume that the shock, once it leaves the ST stage,  
remains radiative, which might not necessarily be the case for the  
low temperatures ($T< 10^5\, \rm{K}$) and shock speeds ($<50\, {\rm km s}^{-1}$)
when the shock has swept up most of the ICM, as cooling is less efficient in 
this regime (Sutherland \& Dopita 1993).

Using $E_{51}=3\times10^{-4}$ and $n=20-150, \rm {cm}^{-3}$ we obtain 
$v_{s}\approx13-15\,{\rm km\, s}^{-1}$ after the shock has
swept up $15\, {\rm M}_{\odot}$. Since this velocity is lower than
the escape speed of $\approx40\, {\rm km\, s}^{-1}$ for M15, it 
follows that the energy of the ejected material from one MSS collision may 
not be sufficient to remove the ICM completely. In
fact, only for $E_{51}\gtrsim8\times10^{-4}$, which is at the upper end of
the estimated ejecta energy interval, the shocked 
ICM gas attains a velocity sufficiently large to leave the cluster.

On the other hand, the shock
also heats up the gas to high temperatures,  
($\approx10-15\times10^{4}\,{\rm K}$ at those shock speeds). 
Gas at such temperatures expands beyond
a critical radius and flows out of the cluster as a wind, reducing
the amounts of ionized gas down to less than $1\,\rm{M}_{\odot}$ 
(Knapp et al.\ 1996).
Assuming that the dust follows the gas, the dust would therefore
leave the cluster on a timescale as short as $\sim10^{5}\,{\rm yr}$.
In fact, the wind should be even stronger in our scenario than for the 
static model considered in Knapp et al.\ (1996), since here the gas itself
has a considerable outwards speed through the shock.

So far, we assumed the existence of rather large amounts of gas,  
based on the amount of observed dust and M15's extremely low metallicity.
However, searches for gas in M15 have had very limited success, and 
Smith et al.\ (1995) estimate an upper limit for the total ICM mass of about 
$3\, {\rm M}_{\odot}$. The reason for the
much larger observed dust-to-gas ratio is not well understood
(see, e.g., van Loon et al.\ 2006). As the dust-to-gas ratio of the
material lost in the winds of red giants should scale with the  
metallicity of the stars (van Loon et al.\ 2005), thus resulting in more 
than $10\,{\rm M}_\odot$ of gas, it appears that additional processes may be 
at work that remove the gas more easily than the dust. Nevertheless, it is
also possible that most of the ICM is in molecular form as the
CO-to-H$_2$ conversion factor is not known for such low metallicities
and extreme radiative environments (van Loon et al.\ 2006). Repeating the 
previous calculation for an ICM mass 
of $M=3\, {\rm M}_{\odot}$ and $n=3-30\, \rm{cm}^{-3}$ accordingly, we obtain
$v_{s} = 52-63\,{\rm km\, s}^{-1}$ for $E_{51}= 3\times10^{-4}$, while
for $E_{51}= 2\times10^{-4}$ $v_{s}=39-47\, {\rm km\, s}^{-1}$
which is close to and larger than the cluster escape speed of 
$40 \rm{km\, s}^{-1}$. In this case, the ejecta of one 
MSS collision would likely be able to accelerate this gas 
out of the cluster.

As a final caveat, we note that it is not very clear whether
the dust will follow the rather low-density ionized gas. This
strongly depends, among other quantities, on the dust grain properties
and their electric potential relative to the ionized gas in a rather
complicated way (Draine \& Salpeter 1979). For instance, Nozawa et
al.\ (2006) and similarly Slavin et al.\ (2004) found in their  
simulations of shocks driven into dusty interstellar medium that, 
while small grains with sizes of $\approx0.01\,{\rm \mu m}\,$ get 
destroyed by sputtering and grain-grain collisions, only grains with 
sizes $\approx0.1\,{\rm \mu m}$ are actually dragged along with the gas, 
while grains with sizes $\geq1\, {\rm \mu m}$
remain almost unaffected and do not follow the shock wave. On the
other hand, if the size distribution of the dust grains is similar
to the one for the local interstellar medium, we see (e.g., Mathis 1996)
that most of the dust mass is in grains with sizes of $\approx0.1\, 
\rm{\mu m}$. This means that, even if not all dust particles follow the gas
flow, we can nevertheless expect that most of the dust mass will
remain sufficiently well coupled to the gas and consequently be removed.

\section{Conclusions}
\label{conclusion}

In this Letter we proposed a new mechanism to explain the  
relatively
short lifetimes of the ICM dust in a dense GC, developing our  
arguments in detail
for the case of M15. By calculating the
rate of stellar collisions using the detailed model for M15 by  
McNamara et al.\ (2004),
we find a remarkable, close agreement between the average time between  
collisions
and the inferred dust lifetime of $\simeq10^{6}\,{\rm yr}$ (Boyer et  
al.\ 2006) in this
cluster, pointing to a direct link between the two phenomena. We  
argue that
the kinetic energy of the material ejected during a stellar
collision may be sufficient to remove the dust from the cluster, 
depending on the assumed ICM mass, either directly by  
accelerating
dust and gas to velocities larger than the cluster escape speed, or  
indirectly,
by accelerating and heating the gas, which then expands and leaves
the cluster as a wind, carrying the dust along with it. Although there
are some uncertainties as to how well the dust will couple
to the gas, especially at the low shock speeds expected for this
problem, there are some indications from simulations of supernova
remnants that might support sufficient coupling (e.g., Nozawa et al.\  
2006).
In addition, at least some dust grains can also
be efficiently destroyed by grain-grain collisions or sputtering (Slavin
et al.\ 2004; Nozawa et al.\ 2006), which further helps to reduce the
amount of observable dust in the cluster. On the other hand,
the results of these studies may not be directly applicable to our
scenario because, e.g., the intense UV field
present in a cluster like M15 could strongly affect the electric  
potential of the grains and, therefore, their coupling to the ionized gas
(Draine \& Salpeter 1979).

With a detailed model for M15 we were also able to calculate the
formation rate and mass distribution of BSs through MS-MS collisions. 
By comparing with the observed number of BSs
in the cluster, we derive an upper limit to their average lifetime
of $\simeq1\,{\rm Gyr}$, consistent with current stellar structure  
and evolution
models for BSs (e.g., Leigh et al.\ 2007).

We conclude that the interaction of ejected gas from
stellar collisions with the ICM will strongly influence the  
observable signal of the dust
in the ICM, and, given the remarkable agreement between dust  
lifetimes and collision times
in M15, this represents a promising mechanism to explain the very  
short dust and ICM lifetimes
in GCs.

\acknowledgements{We thank Holger Baumgardt for providing us with
snapshots of his M15 models and for helpful comments.
We also thank James Lombardi for helpful discussions regarding
stellar collisions, Bruce Draine for valuable advice during a
recent visit to Northwestern, and the referee, Jacco van Loon, who
helped us improving the clarity of the paper.
This work was supported by NASA Grants NNG04G176G and NNX08AG66G.}

%\bibliographystyle{/home/stefan/.TeX/bibtex/apj}
%\bibliography{DusBS}

\end{document}